\documentclass[pre,twocolumn,aps,floatfix,10pt]{revtex4}
\usepackage{amsmath}
\usepackage{amsfonts}
\usepackage{amssymb}
\usepackage{graphicx}
\usepackage{fontenc}
\usepackage{psfrag,layout}
\usepackage{upgreek}

\begin{document}
\title{Excess free energy of supercooled liquids at disordered walls }
\author{Ronald Benjamin and J{\"u}rgen Horbach}
\affiliation{Institut f{\"u}r Theoretische Physik II, Universit{\"a}t D{\"u}sseldorf,
Universit\"atsstra\ss e 1, 40225 D{\"u}sseldorf, Germany}

\begin{abstract}
Using a novel thermodynamic integration scheme, we compute the excess
free energy, $\gamma$, of a glass-forming, binary Lennard-Jones liquid in contact
with a frozen amorphous wall, formed by particles frozen into a similar
structure as the liquid. We find that $\gamma$ is non-zero, becoming
negative at low temperature. This indicates that the thermodynamics of the
system is perturbed by the effect of the amorphous wall.
\end{abstract}

\maketitle

Recently, several studies have investigated the
relaxation dynamics of glass-forming systems in contact with
a wall, formed by particles that are frozen into a similar disordered
structure~\cite{krackoviakjcp,cavagna2009,scheidler_epl2002,scheidler2004,kobnaturephys,berthier2012,cavagna2012,gradenigo2013,cammarota2012}.
One of the objectives in these studies has been to identify a growing
static length scale associated with the dramatic slowing down of glassy
dynamics in the bulk~\cite{ktw1989}. This is based on the assumption that
the thermodynamics of the system is unaffected by the disordered wall,
provided an average is performed over the thermal fluctuations as well
as a sufficient number of wall realizations~\cite{berthier2012}.

Since in such systems, the mobile particles are in contact with an
amorphous wall, albeit whose structure is similar to that of the
liquid, it involves the presence of an interface. A key question
then is to ask what the free energy cost of the formation of such an
interface is.  The absolute free energy of a system is difficult to compute
directly, as the free energy is not a simple function of the phase space
variables. However, in an atomistic simulation the free-energy difference
of a given state from a reference state of known free energy can be
computed using thermodynamic integration (TI)~\cite{frenkel-smit02}.

In this work, we compute the excess free energy, $\gamma$, of a binary
glass-forming Lennard-Jones (LJ) liquid~\cite{kob-andersen-1995} in contact
with quenched disordered walls on either side over the bulk free energy,
using a novel TI scheme in combination with molecular dynamics (MD)
simulation~\cite{allen-tildesley87}. We consider the distance between the
walls to be large enough for the two interfaces to be independent of each
other. The interfacial free energy $\gamma$ is computed as a function of
temperature. For low temperatures, we find that $\gamma$ becomes negative
and thus the amorphous wall imposes an attractive pinning field on the
supercooled liquid.  Furthermore, the non-zero value of $\gamma$ indicates
that the free energy of the liquid is affected by the amorphous walls,
although these walls have a similar structure as that of the liquid.
Therefore, one has to be careful with respect to the interpretation of the
relaxation behavior of the supercooled liquid and following conclusions
about the structural relaxation in the bulk liquid.

{\it {\underline{Model Potential.-}}} We consider a two-component
(particles of type A and B) 80:20 Kob-Andersen (KA) binary Lennard-Jones
mixture~\cite{kob-andersen-1995} with the interaction parameters
chosen to yield a supercooled liquid at low temperatures.  We denote
the interaction potential by $u(r)$, $r$ being the distance between
the particles.  The $N$ binary LJ particles are enclosed within
a simulation cell of size $L_{\text{x}}\times L_{\text{y}} \times
L_{\text{z}}$, with periodic boundary conditions in the $x$ and $y$
directions. Along the z direction, the particles
are confined by disordered walls such that there are
two wall-supercooled liquid interfaces with each interface
having an area $A=L_{\text{x}}L_{\text{y}}$. In order
to prevent the supercooled liquid from penetrating the disordered wall,
a short-ranged flat wall, modelled by a shifted and scaled WCA potential
is introduced, similar to the hard wall introduced by Scheidler {\it
et al.}~\cite{scheidler_epl2002,scheidler2004}.

To integrate the equations of motion, the velocity form of
the Verlet algorithm was used with a time step $\Delta t=0.002
\sqrt{m\sigma_{AA}^{2}/\epsilon_{AA}}$, where the mass $m$ of all
particles is identical and $\sigma_{AA}$ and $\epsilon_{AA}$ are the
size and interaction strength parameters corresponding to component A.
To maintain constant temperature, the velocity of the particles was
drawn from a Maxwell-Boltzmann distribution at the desired temperature,
every $100$ time steps.

We carried out Molecular Dynamics simulations in the $NVT$ ensemble at
temperatures $T=0.53, 0.75, 1.0, 1.5$ and $5.0$, in reduced units. At
$T=0.53$, the temperature is low enough for the mixture to be in
the supercooled regime and yet the equilibration to take place in
a reasonable simulation time.  At all temperatures, between $5-20$
independent configuration of the system at that particular temperature
were equilibrated, to pick reference configurations for the pinned
disordered wall. To generate the reference configurations, the system
was equilibrated for about $2-3$ times the equilibration times reported
in Ref.~\cite{kob-andersen-1995}.

The disordered wall is constructed by choosing particle positions from a
slab of width $5 \sigma_{AA}$, in the middle of the simulations box, from
an equilibrium configuration. New particles fixed at these instantaneous
equlibrated positions were then juxtaposed on the left and right sides
of the simulation cell from the two halves of the slab each of width
$2.5\sigma_{AA}$ [equal to the cut-off range of the interaction potential
$u(r)$], by shifting their $z$-positions by $-L_{\rm z}/2$ and $L_{\rm
z}/2$, respectively.

\begin{figure}[tba]
\includegraphics[width=3.0in]{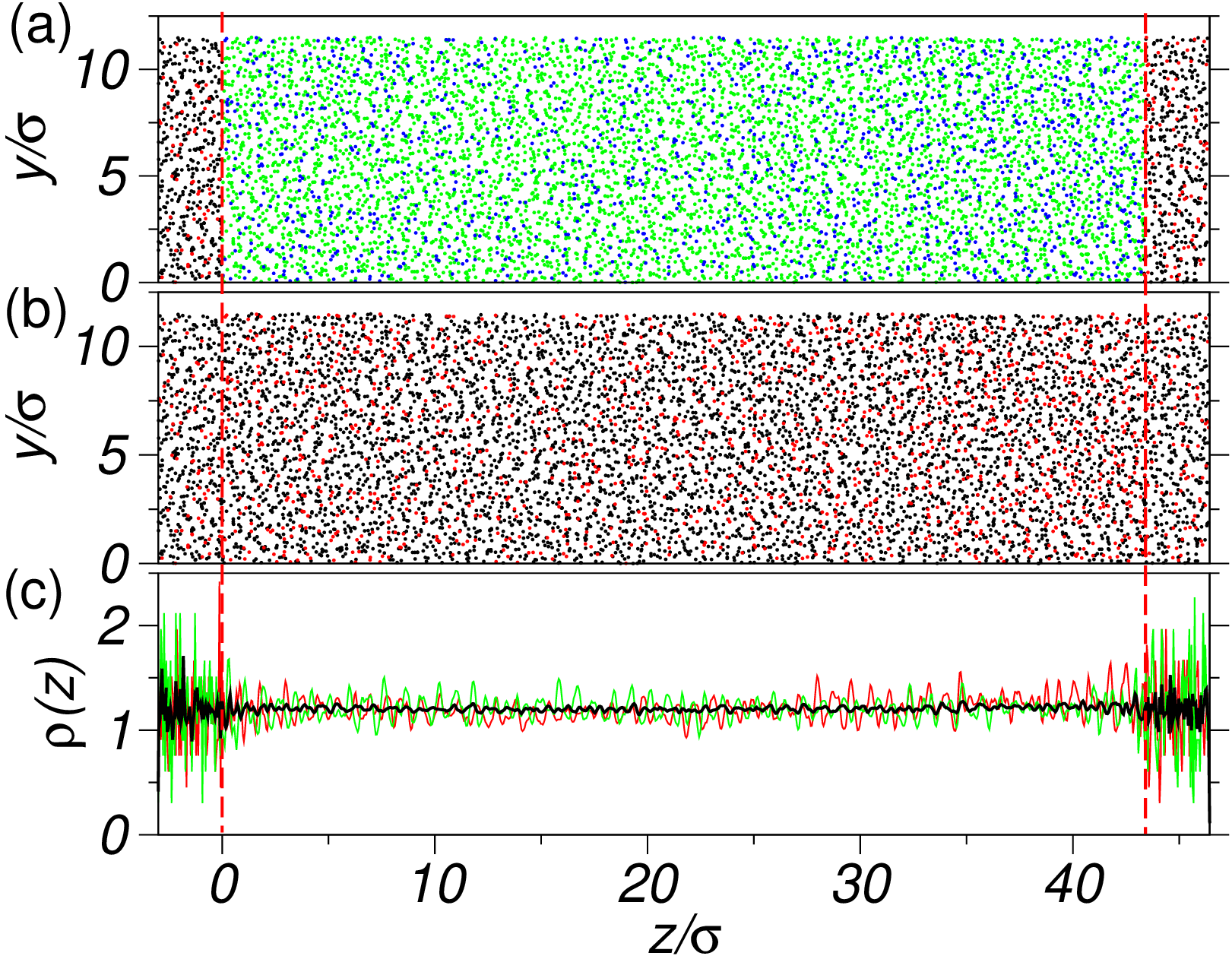}
\caption{\label{fig:profile}(Color online) Simulation setup of supercooled
liquid in contact with disordered walls along the $z$ direction, showing
the two components of the binary liquid mixture by different colors, with
different (a) and same (b) color codes adopted for the mobile and pinned
particles. The dashed vertical line, denotes the location of the flat
walls. (c) Density profile averaged over 10 different wall configurations
and those corresponding to two independent realizations. Data corresponds
to $T=0.53$ and $N=6912$.}
\end{figure}

{\it {\underline{Method.-}}} We adopt a thermodynamic integration scheme
based on a similar approach developed by us earlier to compute the
interfacial free energy of a LJ liquid/crystal in contact with
structured walls~\cite{benjamin-horbach2012} and a method to compute the
crystal-liquid interfacial free energy~\cite{benjamin-horbach2014}. Our
scheme consists of two steps. In the first step, the bulk supercooled
liquid with periodic boundary conditions, is transformed into a state
where it is in contact with short-ranged flat walls on either side
in the $z$ direction, with the periodic boundary conditions still
intact. The structureless flat wall (fw) is described by a shifted
WCA potential and is denoted by, $u_{\text{fw}}(z^\prime)$, where
$z^\prime=z+1.112\sigma_{AA}$. The two flat walls are located at the two
ends of the simulation box along the $z$-direction, at $0$ and $L_{\rm z}$
respectively (for visualization see Fig.~\ref{fig:profile}).  Since the
structureless flat wall has a very short-range ($~0.01\sigma_{AA}$),
the contribution of this step to the total interfacial free energy is
negligible and does not affect the structure of the supercooled liquid
near the disordered walls.

In the second step, the interactions of the bulk system through the
periodic boundaries are gradually switched off while interactions with
the disordered wall are slowly switched on, such that at the end of
the second step we have a supercooled liquid in contact with frozen
amorphous walls on either side in the presence of short-ranged flat
walls. From the free energy differences $\Delta F_1$ and $\Delta F_2$,
corresponding to the two steps, the required interfacial free energy,
$\gamma$, is obtained as, $\gamma=({\Delta F_1 + \Delta F_2})/{A}$.

In both steps of our TI scheme, the transformation from one
equilibrium state to another is carried out by gradually changing
a switching  parameter $\lambda$, which couples directly to the
interaction potential. The $\lambda$ dependent Hamiltonian for
step one is given by: $H_{1}(\lambda)= \lambda^{2}\epsilon_{w}u_{\rm
fw}(z^\prime)$, where $\epsilon_{w}$ varies from $3800\epsilon_{AA}-40000
\epsilon_{AA}$ from the lowest to the highest temperatures considered.
For step two, the $\lambda$ dependent Hamiltonian takes the form:
$H_{2}(\lambda)=(1-\lambda)^{3}u^{\ast}(r)+ \lambda^{3}u_{w}(r+(1-\lambda)
r_{c})$, where $u^{\ast}(r)$ represents interaction between the mobile
particles through the periodic boundaries, while $u_w$ represents
interaction between a mobile and a wall particle and is of the same kind
as the interaction between the free particles, viz.~the 80:20 binary
LJ mixture~\cite{kob-andersen-1995}.

To study finite size effects in the supercooled regime ($T=0.53$),
we considered three system sizes with $N=4000$, $6912$, and $13500$
particles having dimensions $10.0 \times 10.0 \times 33.22$,  $11.5 \times
11.5 \times 43.41$, and $13.0 \times 13.0 \times 66.35$, respectively (in
units of $\sigma_{\rm AA}^{3}$). At $T=0.53$, data was averaged over $20$
independent wall configurations on either side of the liquid for $N=4000$
particles, while for the larger system sizes, $10$ realizations were
considered.  At the higher temperatures, only $5$  independent
wall and liquid configurations were considered and with a system size of
$N=4000$ particles.

\begin{figure}
\includegraphics[width=3.0in]{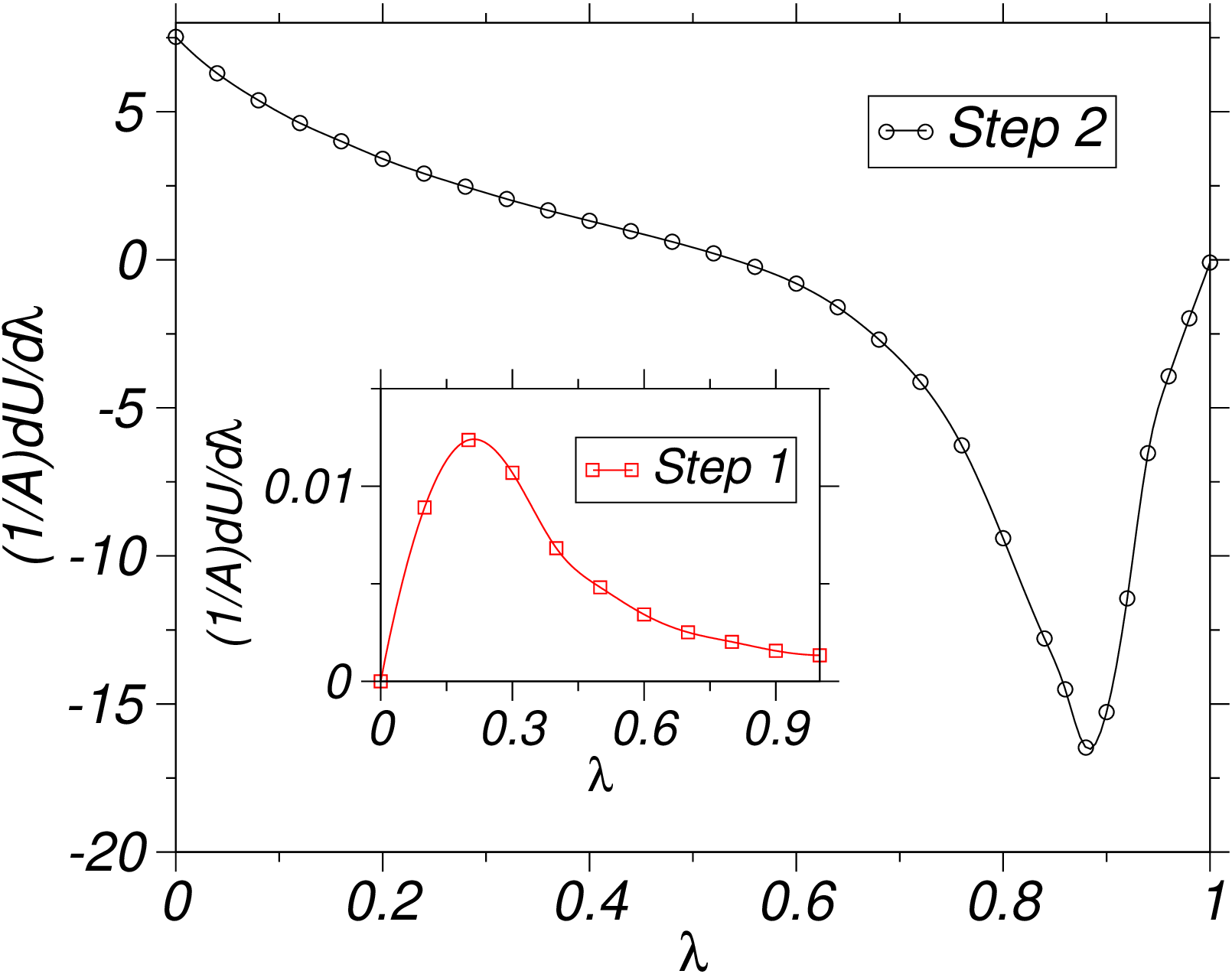}
\caption{\label{fig:step12_dudlambda}(Color online) Thermodynamic
integrand as a function of $\lambda$ for the first (in the inset) and
second steps of the TI scheme at the temperature $T=0.53$ and $N=4000$.}
\end{figure}
\begin{figure}
\includegraphics[width=3.0in]{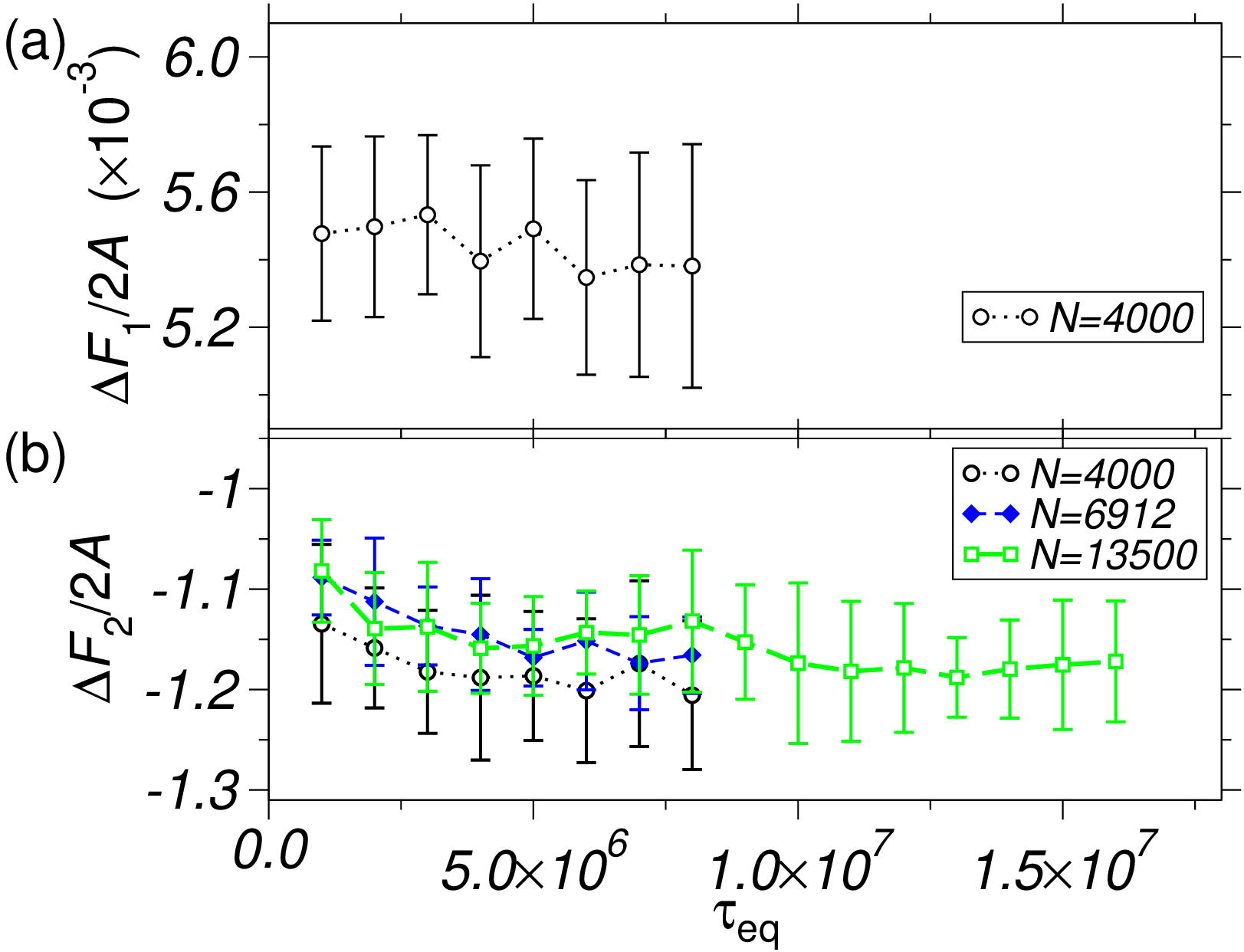}
\caption{\label{fig:step12_gamma}(Color online) Interfacial free energy
as a function of the equilibration time corresponding to the first (a)
and second (b) TI steps for $T=0.53$ and $N=4000$.}
\end{figure}
\begin{figure}
\includegraphics[width=3.0in]{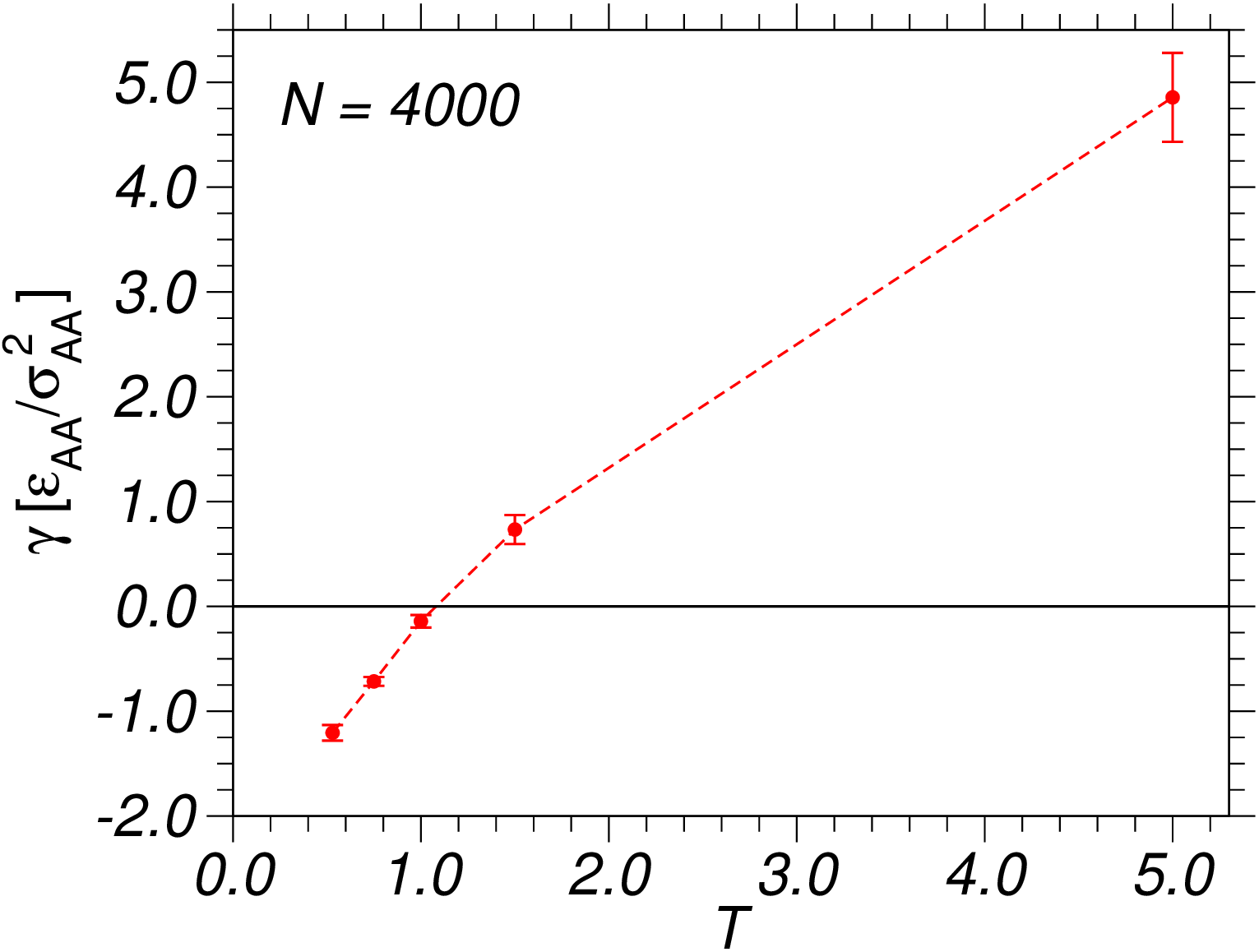}
\caption{\label{fig:total_gamma}(Color online) Interfacial free energy
vs.~temperature for supercooled liquid in contact with a disordered wall
from the sum of the two TI steps for $N=4000$.}
\end{figure}

{\it{\underline{Results.-}}} The simulation set-up for the supercooled
liquid in contact with a wall made of frozen-in particles is shown
in Fig.~\ref{fig:profile}, corresponding to the temperature $T=0.53$.
As the two snapshots in Fig.~\ref{fig:profile} indicate, there
is no change in the structure of the liquid near the walls and
representing all regions by the same color scheme one is not able
to distinguish any interface (Fig.~\ref{fig:profile}b). This is also
reflected in the density profile averaged over $20$ independent wall
configurations, with the profiles corresponding to two individual runs
(Fig.~\ref{fig:profile}c). Averaging over the thermal fluctuations and
independent realizations of the disordered wall, leads to a flat density
profile near the wall as in the bulk. Hence the pinned particles do not
affect the structural behavior of the free particles in the vicinity of
the wall. We have also observed a similar behavior corresponding to the
potential energy profile, in agreement with the findings by Scheidler
{\it et al.}~\cite{scheidler2004}.

While this might lead to the conclusion that thermodynamics of the free
particles is unperturbed in the presence of such a quenched disordered
wall, in the following we show by our TI scheme that a non-zero excess
free energy is associated with the interface between the wall and the
liquid.  The thermodynamic integrands corresponding to the two steps
of the TI calculation are displayed in Fig.~\ref{fig:step12_dudlambda},
corresponding to the temperature $T=0.53$ and the system with $N=4000$
particles. At this temperature, the time scale for structural relaxation
near the wall increases by several orders of magnitude in comparison
to the bulk~\cite{scheidler2004}.  The data reported in this figure
correspond to an averaging over the $20$ different wall configurations
and the thermal fluctuations. At the various independent wall
realizations, the system was equilibrated for $\tau_{\rm eq} = 8\times 10^{6}$ 
time steps at each value of $\lambda$. Then production runs were carried
out for $50000$ time steps during which the averages were calculated.

We observe in Fig.~\ref{fig:step12_dudlambda} that the thermodynamic 
integrand at various values of the $\lambda$ parameter, corresponding 
to step one is much smaller in magnitude than the corresponding curve 
for step two. Also, as a non-linear parameterization is adopted for 
the varying the interaction potentials, the thermodynamic integrands 
are not monotonically varying. In step two, initially the interaction 
with the disordered wall is weak while the periodic boundary conditions 
are being gradually switched off, resulting in positive values for the 
thermodynamic integrand. When the interaction with the disordered wall 
becomes stronger, the curve becomes negative and varies in a non-monotonic 
manner as $\lambda$ goes to one.

In order to check whether $\tau_{\rm eq}$ is sufficient for
the equilibration of the system at each value of $\lambda$, we
have varied $\tau_{\rm eq}$ from $\tau_{\rm eq}=1\times10^6$ to
$\tau_{\rm eq}=16\times10^6$ and computed the free energy differences
corresponding to the two steps as a function of $\tau_{\rm eq}$. For
the case of $\Delta F_1/(2A)$ (Fig.~\ref{fig:step12_gamma}a), only
small variations are seen as a function of $\tau_{\rm eq}$ and $\Delta
F_1$ saturates to a value of $0.0054(3) \epsilon_{AA}/\sigma_{AA}^2$
(the number in parenthesis indicates the error in the last digit).
While the data shown in (Fig.~\ref{fig:step12_gamma}a) corresponds to
$N=4000$ particles, the results for the larger systems (only computed
for $\tau_{\rm eq}=8\times10^6$) are also in good agreement with this
value of $\Delta F_1/(2A)$.

For the calculation of $\Delta F_2/(2A)$, we varied the time $\tau_{\rm
eq}$  up to $16\times 10^{6}$ time steps for $N=13500$ particles and
up to $8\times 10^{6}$ time steps for $N=4000$ and $6912$ particles. The
corresponding data is displayed in Fig.~\ref{fig:step12_gamma}b. For
all system sizes, $\Delta F_2/(2A)$ at first slightly decreases as
a function of $\tau_{\rm eq}$ and then saturates to an average
value around $\Delta F_2/(2A)=-1.206(74)\epsilon_{AA}/\sigma_{AA}^2$ after 
about $4\times10^{6}$ time steps, for the three system
sizes (in fact, within the error bars, finite-size effects are minor).  
The comparison of $\Delta F_1$ and $\Delta F_2$ indicates that
contribution of $\Delta F_1$ to total free energy is negligible. 
At the higher temperatures also, $\Delta F_1 \ll  \Delta F_2$.

In Fig.~\ref{fig:total_gamma}, we show the interfacial free energy
$\gamma$ as a function of temperature, corresponding to the system
size $N=4000$. At all temperatures reported in this work, the potential
energy per particle is always negative and the structure of the liquid
is the same near the vicinity of the wall as in the bulk. The free
energy difference $\Delta F=\Delta U - T \Delta S$, where $\Delta U$
is the change in the internal energy and  $\Delta S$ is the change
in the entropy of the system. Clearly, since $\Delta U=0$, a non-zero
free energy difference can only arise due to a change in the entropy
of the system. Since the interfacial free energy at $T=0.53$ turns out
to be negative at this temperature, the entropy of the liquid in the
supercooled regime increases in presence of such a disordered wall.
With increasing temperature, however, $\gamma$ increases and eventually
becomes positive. At higher temperatures, therefore the change in entropy
is negative.

From these findings, one can infer that such a disordered wall exerts
an effective pinning field on the particles, clearly perturbing the
thermodynamics even though the structure of the liquid near the wall
remains unchanged.  At low temperatures, the negative value of $\gamma$
indicates that the disordered wall exerts an attractive potential on
the confined supercooled particles acting as a trap, thus slowing down
the dynamics. As temperature increases this potential field becomes
positive. It is plausible, therefore, that this behavior of $\gamma$
with respect to the temperature, is correlated to the relaxation
time of the supercooled liquid near the wall, which as previous works
have shown, decreases in the neighborhood of the wall as temperature
increases~\cite{scheidler_epl2002}. It is to be noted that for 
glass-forming systems strongly confined by parallel flat walls, a 
Mode-coupling theory has been developed recently to explain the
slowing down of the relaxation dynamics~\cite{lang2009}. However, no 
such microscopic theories exist for supercooled liquids in contact 
with disordered walls.

While here we have studied the effect of pinned particles on a
supercooled liquid, similar behavior of the interfacial free energy is
also observed in other systems. In a previous work, as part of a TI scheme
to compute the crystal-liquid interfacial free energy, we computed the
free energy difference between a bulk crystal and the crystal in contact
with a structured wall made of crystalline particles frozen into their
equilibrium positions~\cite{benjamin-horbach2014}. A  negative
value for the interfacial free energy was obtained (see Fig.~4 of 
Ref.~\cite{benjamin-horbach2014}, where the area in the negative region 
under the thermodynamic integrand curve is much greater than the 
positive part) even though the structure and potential
energy of the crystal in contact with such a wall is the same as in
the bulk. We also observed that the free energy difference became less
negative at higher temperatures. This shows that even for the simpler case
of a crystal evolving under the effect of pinned crystalline particles,
the thermodynamics of the moving particles is definitely perturbed by
the presence of the pinned particles and is not the same as in the bulk.

{\it{\underline{Conclusion.-}}} In this work, we have obtained the interfacial
free energy of a supercooled liquid in contact with frozen amorphous
walls having the same structure as the liquid. We obtain a negative value
for the interfacial free energy at low temperatures which becomes less
and less negative at higher temperatures and then reaches a positive
value. This indicates that thermodynamics of the supercooled liquid is
not the same as in the bulk but is perturbed by the presence of such
a disordered wall. Since the structure of the liquid in the vicinity
of the wall is the same as in the bulk, the non-zero interfacial free
energy is a purely entropic effect.

Our results have implications for studies which interpret the relaxation
dynamics of supercooled liquids in the neighborhood of an amorphous wall
to be correlated to the bulk behavior since such studies are based on the
assumption that the pinned particles do not affect the thermodynamics
of the confined liquid. However, one has to be careful when drawing
conclusions about the bulk behavior of a supercooled liquid from the
dynamics of the liquid near an amorphous wall, since the thermodynamics
of the system is affected by the pinned particles.

\begin{acknowledgments}
The authors acknowledge the research unit FOR 1394 ``Nonlinear response 
to probe vitrification'' of the Deutsche Forschungsgemeinschaft (DFG),
Grant No.~HO 2231/7-2.
\end{acknowledgments}

\end{document}